\newcommand{\be}{\begin{equation}}
\newcommand{\ee}{\end{equation}}

\documentclass{mn2e}
\input{epsf}

\title[Tidal tails around dwarf galaxies]
{The orientation and kinematics of inner tidal tails around dwarf galaxies orbiting the Milky Way}

\author[J. Klimentowski et al.]
    {Jaros{\l}aw Klimentowski,$^{1}$ Ewa L. {\L}okas,$^{1}$ Stelios Kazantzidis,$^{2}$
    Lucio Mayer,$^{3,4}$ \newauthor{Gary A. Mamon}$^{5,6}$ and Francisco Prada$^{7}$
    \\
    \\
    $^1$Nicolaus Copernicus Astronomical Center, Bartycka 18,
    00-716 Warsaw, Poland\\
    $^2$Center for Cosmology and Astro-Particle Physics;
        and Department of Physics; and Department of Astronomy, \\
    The Ohio State University, Physics Research Building, 191 West Woodruff Avenue, Columbus, OH 43210, USA\\
    $^3$Institute for Theoretical Physics, University of Z\"urich, CH-8057 Z\"urich, Switzerland\\
    $^4$Institute of Astronomy, Department of Physics, ETH Z\"urich, Wolfgang-Pauli
    Strasse, CH-8093 Z\"urich, Switzerland  \\
    $^5$Institut d'Astrophysique de Paris (UMR 7095: CNRS and Universit\'e Pierre \& Marie Curie),
    98 bis Bd Arago, F-75014 Paris, France \\
    $^6$Astrophysics \& BIPAC, University of Oxford, Keble Rd, Oxford OX1 3RH, United Kingdom \\
    $^7$Instituto de Astrof{\'\i}sica de Andalucia (CSIC), Apartado Correos 3005, E-18080 Granada, Spain
    }
\begin{document}

\maketitle

\begin{abstract}
Using high-resolution collisionless $N$-body simulations we study the
properties of tidal tails formed in the immediate vicinity of a two-component dwarf galaxy evolving in a
static potential of the Milky Way (MW). The stellar component of the dwarf is initially in the form of a disk and the
galaxy is placed on an eccentric orbit motivated by CDM-based cosmological simulations.
We measure the orientation, density and velocity distribution of the stars in
the tails. Due to the geometry of the orbit, in the vicinity of the dwarf, where
the tails are densest and therefore most likely to be detectable, they are typically oriented
towards the MW and not along the orbit. We report on an interesting phenomenon of `tidal tail flipping': on the way
from the pericentre to the apocentre the old tails following the orbit are dissolved and new ones pointing towards the
MW are formed over a short timescale. We also find a tight linear relation between the velocity of stars in the tidal
tails and their distance from the dwarf. Using mock data sets we demonstrate that if dwarf spheroidal (dSph) galaxies
in the vicinity of the MW are tidally affected their kinematic samples are very likely
contaminated by tidally stripped stars which tend to artificially inflate the measured velocity dispersion. The effect
is stronger for dwarfs on their way from the peri- to the apocentre due to the formation of new tidal tails after
pericentre. Realistic mass estimates of dSph galaxies thus require removal of these stars from kinematic samples.

\end{abstract}


\section{Introduction}

The population of dwarf galaxies in the Local Group offers a unique possibility to test current models of structure
formation in the Universe based on $\Lambda$CDM cosmology. Among those, the dwarf spheroidal (dSph) galaxies have
attracted most attention, as they are believed to be strongly dark matter dominated. Many formation scenarios of these
objects have been proposed, including gravitational processes such as tidal interactions with the host galaxy (Mayer
et al. 2001; Kravtsov, Gnedin \& Klypin 2004; Kazantzidis et al. 2004) and those inherent to the baryonic component
such as cooling, star formation, feedback from supernovae and UV background radiation that can occur also for galaxies
in isolation (Ricotti \& Gnedin 2005; Mayer et al. 2007; Tassis, Kravtsov \& Gnedin 2008; Sawala, Scannapieco \& White
2009).

According to the tidal stirring scenario (Mayer et al. 2001) dSph galaxies of the Local Group formed
via tidal interaction of low surface brightness disky dwarfs with the host galaxy such as Milky Way (MW). Inherent in
this process is a significant mass loss which accompanies the morphological transformation.
Matter stripped from the dwarf galaxy forms tidal tails which will eventually become completely
unbound and form larger tidal extensions tracing approximately the original orbit of the dwarf in the MW halo. The
dynamics of the tidal debris in the immediate vicinity of the dwarf turns out to be different from the behaviour of the
debris at larger scales. In this paper we concentrate on the properties of the inner tidal tails, within a few
characteristic radii of the dwarf which are the most observable, given their high surface density.
This debris has been lost recently and remains close to the dwarf and is thus
most likely to contaminate the kinematic samples of dSph galaxies.

The presence of tidal tails around observed dSph galaxies has been detected in many cases.
Mart{\'\i}nez-Delgado et al. (2001) found tidal extensions in Ursa Minor and
Coleman et al. (2005) showed differences in surface brightness around Fornax
depending on the direction of measurement. Tidally induced overdensities in the distribution
of stars were also detected around Carina (Mu\~noz et al. 2006) and Leo I (Sohn et al. 2007). However, owing to their
low surface brightness, tidal tails are still very difficult to detect by photometric techniques. Another possibility
is to look for velocity gradients that should be present in kinematic data sets. Such features seem to be present in
Leo I where the intrinsic rotation appears reversed further away from the centre of the dwarf ({\L}okas et al. 2008).
Note however that the features usually interpreted as tidal could also originate from multiple stellar populations
(Tolstoy et al. 2004; McConnachie, Pe\~narrubia Jorge \& Navarro 2007) or accretion and merger events (Coleman et al.
2004; Battaglia et al. 2006).

It has long been debated whether the inferred large dark matter content of dSph galaxies can in fact be due to the tidal
effects and models with no dark matter were constructed (e.g. Kroupa 1997). In our previous work we showed
(Klimentowski et al. 2007, 2009) that tidal stripping naturally coexists for a long time with a substantial bound
stellar component embedded in a relatively massive dark matter halo. We have also demonstrated that stars originating
from tidal tails can contaminate kinematic samples of dSph galaxies leading to significant overestimates of their
velocity dispersions, especially at larger projected radii. Modelling of such contaminated velocity dispersion profiles
results in biased inferences concerning the dark matter distribution and the anisotropy of stellar orbits, i.e. more
extended dark matter haloes or more tangential orbits are obtained (see the discussion of the case of Fornax in
Klimentowski et al. 2007 and {\L}okas 2009).

The degree of this contamination depends however critically on the
orientation of the tails with respect to the line of sight of the observer, which is approximately the direction
towards the MW. A question thus arises if there is a correlation between their orientation and the orbit of the dwarf
galaxy. If such a correlation exists it would also help to study dSph orbits and serve as a complement to proper
motions measurements which still have rather large errors (e.g. Dinescu et al. 2004; Piatek et al. 2007). Such studies
were already made for example by Piatek \& Pryor (1995) and Johnston, Choi \& Guhathakurta (2002).

In this paper we study the orientation of the tails and their kinematic properties using high resolution $N$-body
simulations of a two-component dwarf galaxy described in Klimentowski et al. (2007, 2009). Although the properties of
the tails depend in general on a number of factors, such as the host potential (Piatek \& Pryor 1995) or the mass of
the satellite (Choi, Weinberg \& Katz 2007) we believe that the orbit is the crucial factor. The potential of the MW is
rather well determined and, as we will show, the tails behave similarly in spite of the fact that during the whole
evolution the mass of the dwarf decreases by more than two orders of magnitude due to tidal stripping.

Although we consider only a single orbit, our choice is strongly motivated by studies of the statistics of subhalo
orbit shapes in the context of $\Lambda$CDM cosmology for cluster-size (Ghigna et al. 1998) and Local Group size haloes
(Diemand, Kuhlen \& Madau 2007). In particular, Diemand et al. find in their Via Lactea simulations that the
distribution of pericentre to apocentre values $r_{\rm p}/r_{\rm a}$ of subhaloes is strongly peaked at about 0.1 with
a median of 0.17 with the range of 0.07-0.40 corresponding to the 68 per cent region. Our orbit, with $r_{\rm p}/r_{\rm
a}=0.2$, is thus very typical and the results should be valid for most dwarf galaxies. We rely on this result because
the orbits of dwarf satellites are at present poorly constrained observationally.

The amount of tidally stripped matter is controlled by the orbit-averaged tidal force which depends on the size of the
orbit and can be quantified by the orbital time (Mayer et al. 2001). The orbital time in our simulations, of 2 Gyr,
is typical for MW satellites (Johnston, Sigurdsson \& Hernquist 1999) and results in relative mass loss rates
of the order of 10-20 percent per Gyr near apocentre (Klimentowski et al. 2009), again in agreement with estimates for
nearby dwarfs by Johnston et al. (1999) who find values between a few up to more than 30 percent per Gyr. Therefore the
density of tidal tails forming in our simulations, and the consequent amount of expected contamination, should be
representative of many MW satellites.

The paper is organized as follows. The simulation details are summarized in the next section. In section 3 we
measure the orientation of the tidal tails with respect to the line of sight of an observer placed at the centre of the
Milky Way and in section 4 describe the densities and velocities of the stars in the tails. Section 5 discusses the
actual contamination of the kinematic samples by unbound stars and its effect on the estimated velocity dispersion.
The discussion follows in section 6.

\section{The simulations}

The simulations used for the present study are described in detail in Klimentowski et al. (2007, 2009). Here we provide
a short summary of the most important parameters. The dwarf progenitor of total mass $M = 4.3 \times 10^9$ M$_{\odot}$
consists of a baryonic disk embedded in a dark matter halo. The initial mass of the disk is $1.5 \times 10^8$
M$_{\odot}$ (it is modelled by $N = 10^6$ particles of mass 149.5 M$_{\odot}$) and its density drops exponentially with
radius in the rotation plane (with the radial scale length $R_{\rm d} = 1.3$ kpc) while its vertical structure is
modelled by isothermal sheets (with the vertical scale height $z_{\rm d} = 0.13$ kpc). The mass of the dark matter halo
is $4.1 \times 10^9$ M$_{\odot}$ (it is modelled by $N = 4 \times 10^6$ particles of mass $1035.7$ M$_{\odot}$) and its
density distribution follows the NFW profile with virial mass $M_{\rm vir} = 3.7 \times 10^9$ M$_{\odot}$ and
concentration $c=15$. The halo is exponentially truncated outside the virial radius $r_{\rm vir} \simeq 40.2$ kpc to
keep the total mass finite.

We performed three simulations with the disk
initially positioned at the inclination of 0$^\circ$, 45$^\circ$ and 90$^\circ$ to the orbital plane. Although the
evolution is slightly different in each case, with stronger mass loss for lower inclinations (see section 4 of
Klimentowski et al. 2009), the orientation does not strongly affect the properties of tidal tails. Since the
results are similar in all cases, here we present only the intermediate, and therefore the most representative, case
of 45$^\circ$ inclination.

The dark matter halo is initially much more extended than the disk, so the mass-to-light
ratio is not constant with radius but increases strongly outside $r=2$ kpc. In the final stage, since the dark halo is
more heavily stripped, the mass-to-light ratio is almost constant and of the order of ten solar units. Our dwarf galaxy
is thus moderately dark matter dominated so the results apply directly to objects like Fornax or Leo I dSph galaxies,
rather than more strongly dark matter dominated objects like Draco.

The dwarf galaxy evolves on an eccentric orbit with apocentre $r_{\rm a}=120$ kpc and pericentre to apocentre ratio of
$r_{\rm p}/r_{\rm a} \approx 0.2$ around the host galaxy. The evolution is followed for 10 Gyr corresponding to five
orbital times. The host galaxy is modelled by a static gravitational potential assumed to have the present-day
properties of the MW as described by mass model A1 of Klypin, Zhao \& Somerville (2002). It consists of a NFW halo with
the virial mass of $M_{\rm vir} = 10^{12}$ M$_{\odot}$ and concentration $c=12$, a stellar disk with mass $M_{\rm D} =
4 \times 10^{10}$ M$_{\odot}$, the scale length $R_{\rm d} = 3.5$ kpc and the scale height $z_{\rm d} = 0.35$ kpc, and
a bulge of mass $M_{\rm b}=0.008 M_{\rm vir}$ and scale-length $a_{\rm b} = 0.2 R_{\rm d}$.

The simulation was performed using the
PKDGRAV $N$-body code (Stadel 2001). The gravitational softening length was 50 pc for stars and 100 pc for dark matter.
We used 200 outputs of the simulation saved at equal snapshots of 0.05 Gyr.

\begin{figure}
\begin{center}
    \leavevmode
    \epsfxsize=7.5cm
    \epsfbox[0 0 370 370]{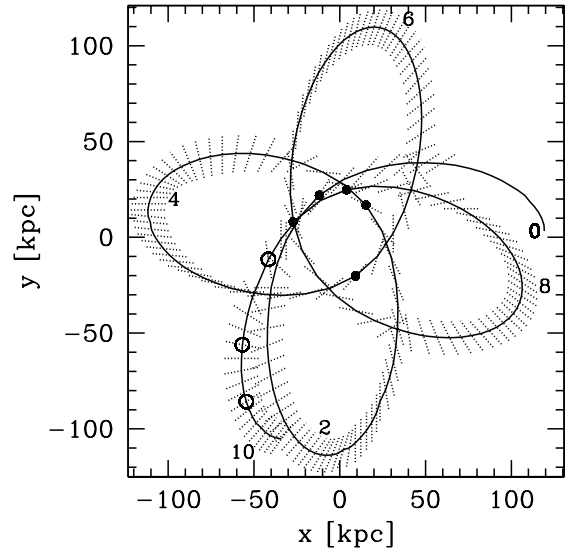}
\end{center}
\caption{The solid line shows the orbit of the simulated dwarf galaxy projected on the plane of
the MW disk. The simulation starts at the right-hand side of the Figure. Numbers marking the apocentres indicate
the time from the start of the simulation in Gyr. Filled circles mark pericentres of the
orbit. Open circles indicate the positions corresponding to the snapshots presented in Fig.~\ref{flipping}. The dotted
lines overplotted on the orbit show the direction of the tidal tails. In the initial part of the orbit
that lacks dotted lines no well defined tidal tails were formed yet.}
\label{orbit}
\end{figure}

\section{The orientation of inner tidal tails}

The evolution of the dwarf galaxy in the potential of the host proceeds via a sequence of tidal shocks when the galaxy
passes the pericentre. The pericentre passages are characterized by significant mass loss so that after 10 Gyr of
evolution the galaxy mass decreases by more than two orders of magnitude. The mass loss is accompanied by the
morphological transformation of the stellar component. Already at the first pericentre passage the disk transforms into
an elongated bar and the prolate shape is preserved until the end, although the galaxy becomes more spherical with
time. At all times pronounced tidal tails can be identified around the dwarf. Here we focus on the innermost tidal
formations, starting near the dwarf, where the stellar density distribution starts to flatten, up to 10 kpc from the
centre of the dwarf. In the following we will refer to this fraction of the tails as `inner tidal tails'.

There are always two tidal tails emanating from the two opposite
sides of the galaxy. While the two tails are not always equilinear (they rarely intersect the centre of the galaxy),
they are always almost parallel to each other so the direction of both tails can be represented by a single straight
line. The alignment of the inner tidal tails in the simulation was calculated by fitting a straight line to the
direction where the density of the tail reaches a maximum. The fitting was done separately for each of the two tails
and the average angle was taken as the final measurement (the two values never differ by more than a few degrees). In
the final output of the simulation the density profile of the stars starts to flatten at about 2 kpc signifying the
transition to tidal tails. Thus our chosen distance of 10 kpc corresponds to about 5 final radii of the dwarf. The
stars in this region are still marginally bound (see below) but are no longer good tracers of the potential of the
dwarf. Except for moments of abrupt changes in the orientation of the tails the estimated alignment extends much
further than 10 kpc.

\begin{figure}
\begin{center}
    \leavevmode
    \epsfxsize=8.5cm
    \epsfbox[10 0 563 183]{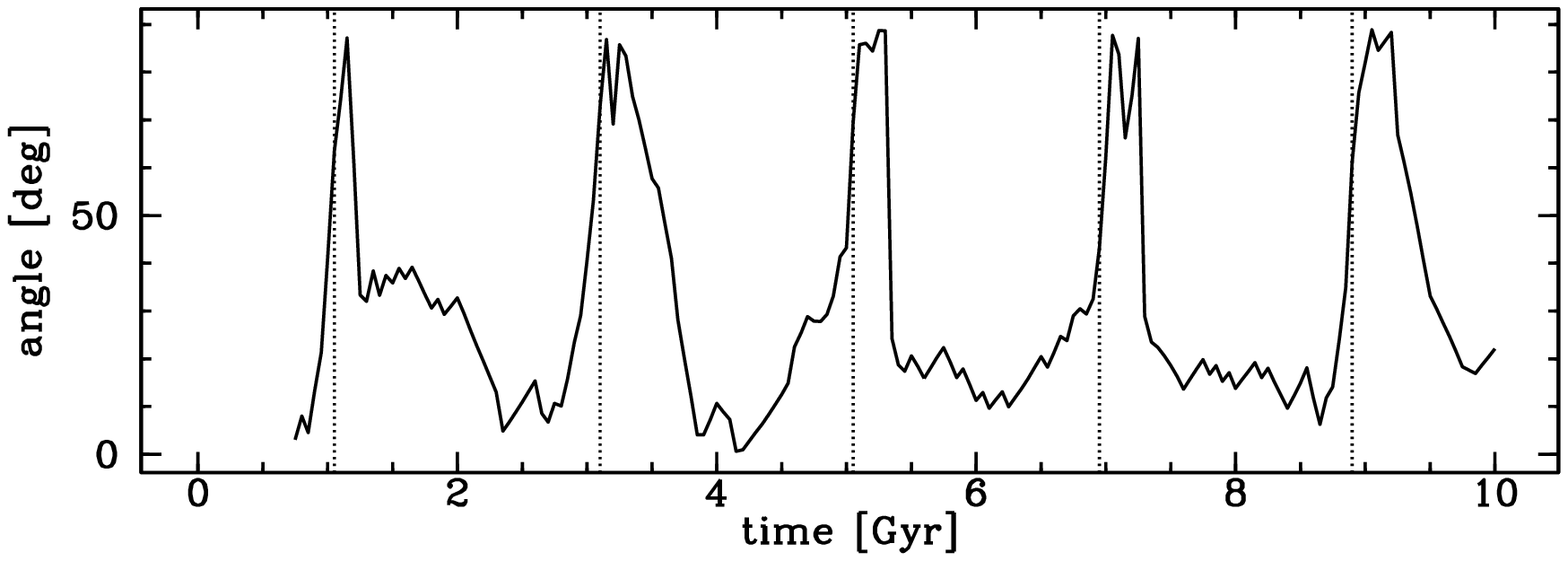}
    \leavevmode
    \epsfxsize=8.5cm
    \epsfbox[10 0 563 183]{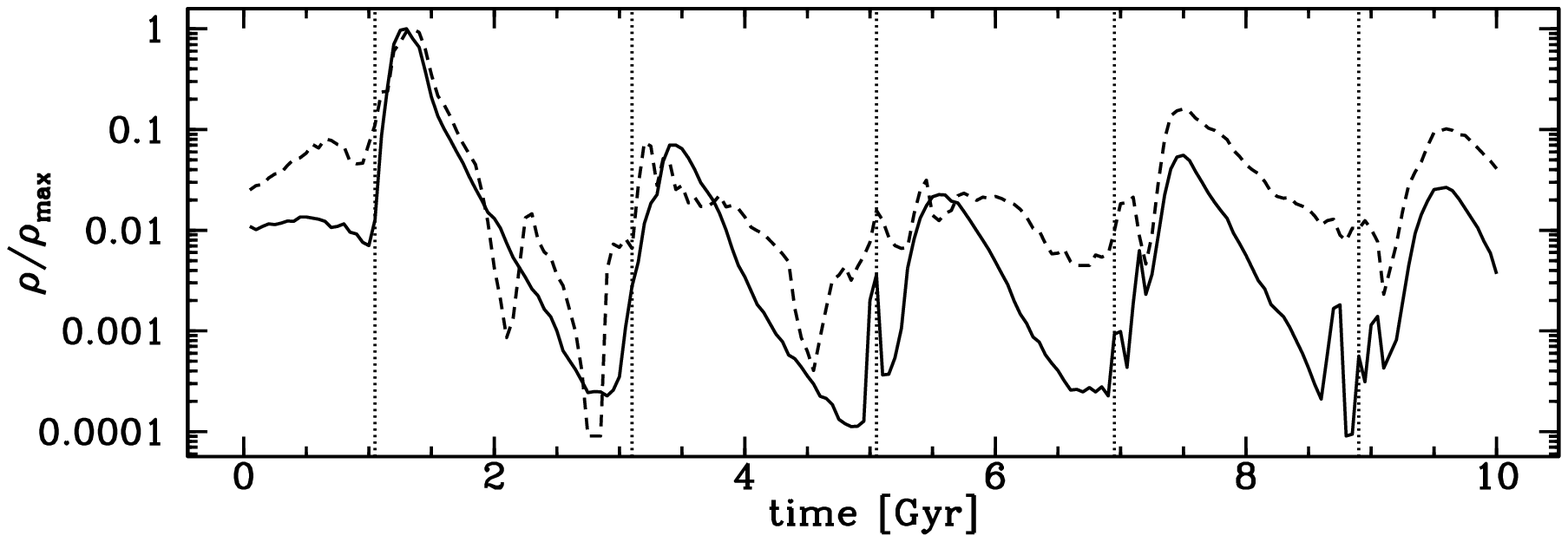}
    \leavevmode
    \epsfxsize=8.5cm
    \epsfbox[10 0 563 183]{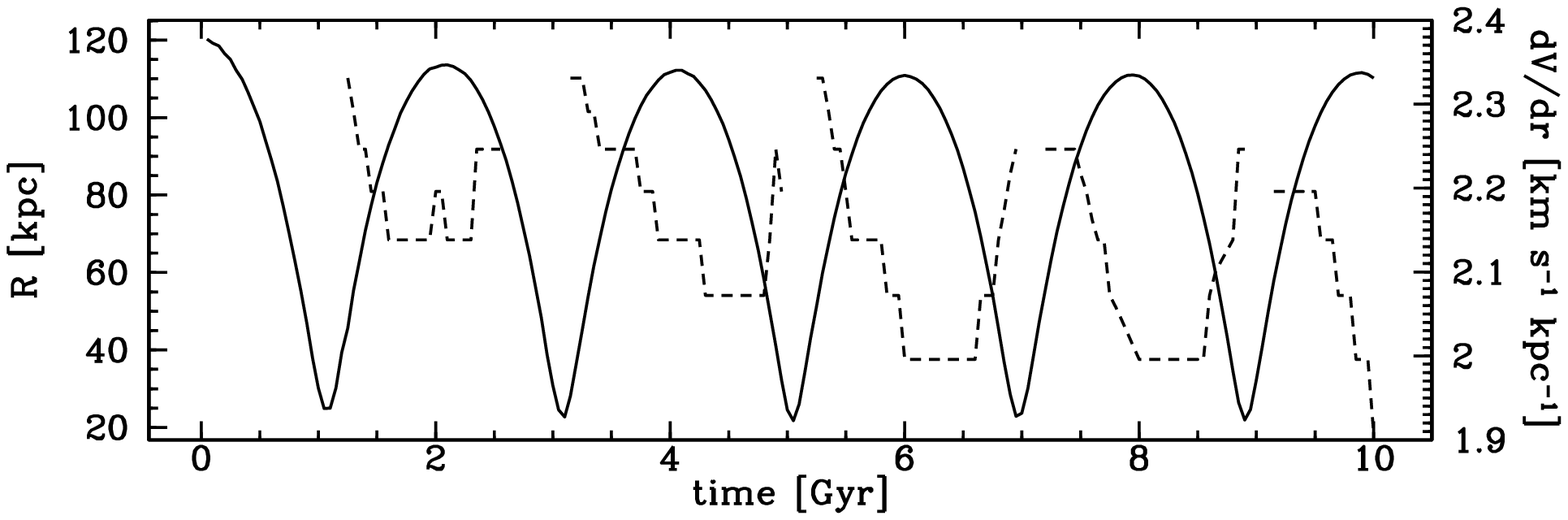}
\end{center}
\caption{Properties of the tidal tails. The upper panel shows the angle between the tails and the direction toward the
centre of the MW as a function of time. The middle panel plots the density of the tidal
tails. The solid line corresponds to the mean density of all stars at the distance of 10 kpc from the dwarf
(calculated in a ring of 1 kpc in size), while the dashed line shows the same density but measured on the axis of the
tails (calculated in two boxes with 1 kpc on a side placed at 10 kpc from the centre of the dwarf in opposite
directions). Densities are plotted relative to their respective maximum values. Vertical dotted lines indicate
pericentre passages. The lower panel shows the values of ${\rm d}V/{\rm d}r$, the slope of the relation between the
velocity of stars in the tidal tails and their distance from the dwarf
(dashed line) fitted to velocity diagrams similar to Fig.~\ref{veltail} along with the distance of the dwarf from the
host galaxy (solid line) as a function of time.}
\label{tails_properties}
\end{figure}

\begin{figure}
\begin{center}
    \leavevmode
    \epsfxsize=8cm
    \epsfbox[26 0 555 280]{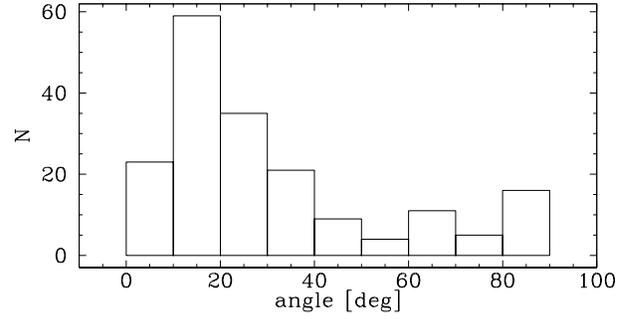}
\end{center}
\caption{The distribution of the angle between the direction to the MW centre and the nearest tidal tail of the
dwarf galaxy. The histogram was constructed from
185 simulation outputs (excluding only the initial 15 outputs where the tidal tails are not yet fully formed).}
\label{ha}
\end{figure}

Fig.~\ref{orbit} plots the orbit of
the simulated dwarf galaxy during the course of its evolution (solid line) with dotted lines showing the direction of
the tidal tails at a given position on the orbit. No
dotted lines are plotted for the initial stage of the simulation when there were no clear tidal tails formed yet.
According to Fig.~\ref{orbit} tidal tails are parallel to the orbit only during very
short periods of time around the pericentres. In the outer parts of the orbit, near the
apocentre (where the dwarf galaxy spends most of its time) tidal tails always point
almost radially towards the host galaxy.
The upper panel of Fig.~\ref{tails_properties} presents the same results in a more
quantitative way by plotting the angle between the tidal tails and the direction to the host galaxy. Tidal tails tend
to be oriented radially: their direction is similar to the direction of the dwarf's position vector for most of the
time. Only for a short time near pericentres they are roughly parallel to the orbit. This result is
illustrated further in Fig.~\ref{ha} where we plot the distribution of this angle
which shows a strong peak for low values of $0^{\circ}-30^{\circ}$.

\begin{figure*}
\begin{center}
    \leavevmode
    \epsfxsize=5.8cm
    \epsfbox[0 0 400 400]{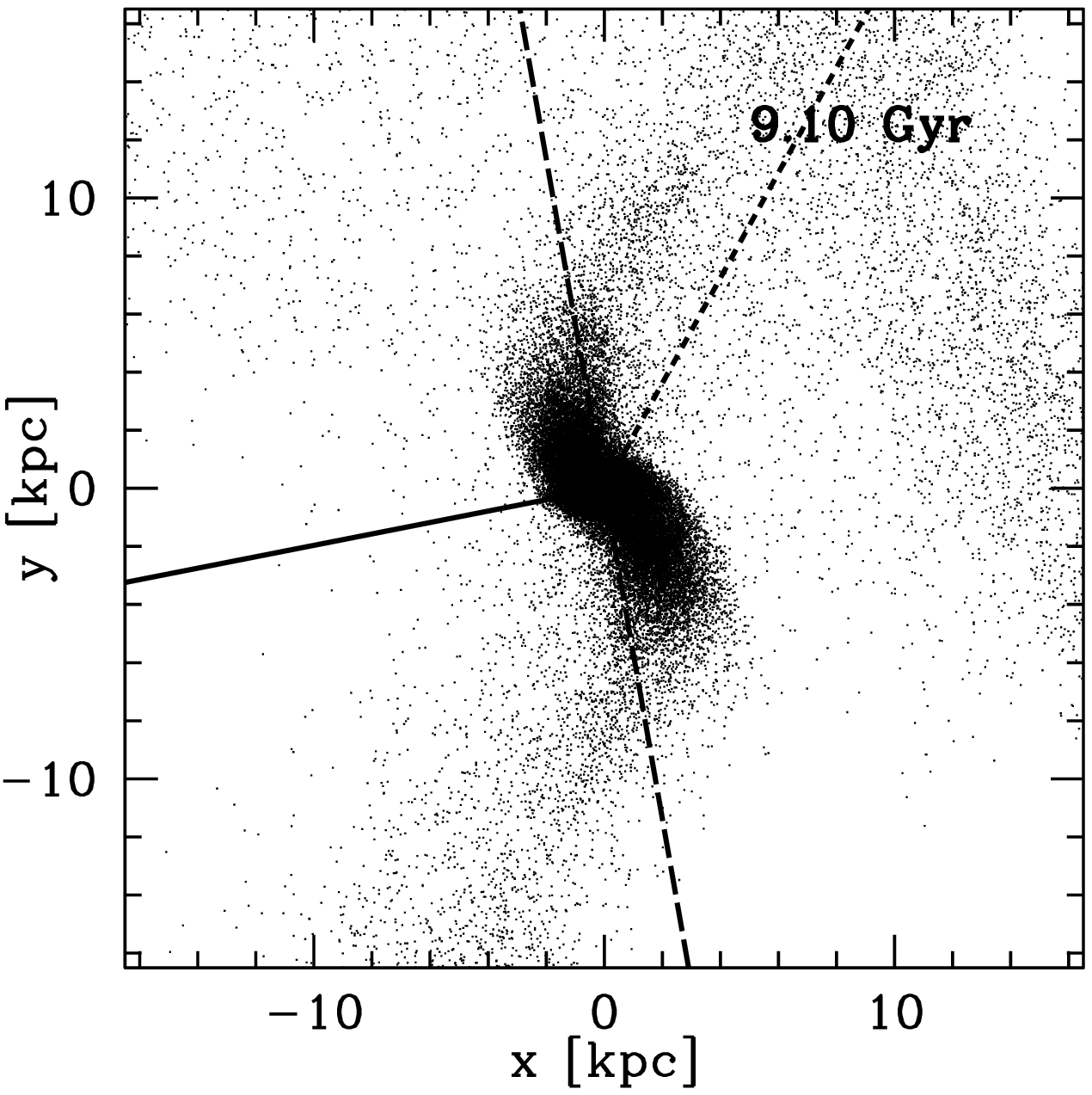}
    \epsfxsize=5.8cm
    \epsfbox[0 0 400 400]{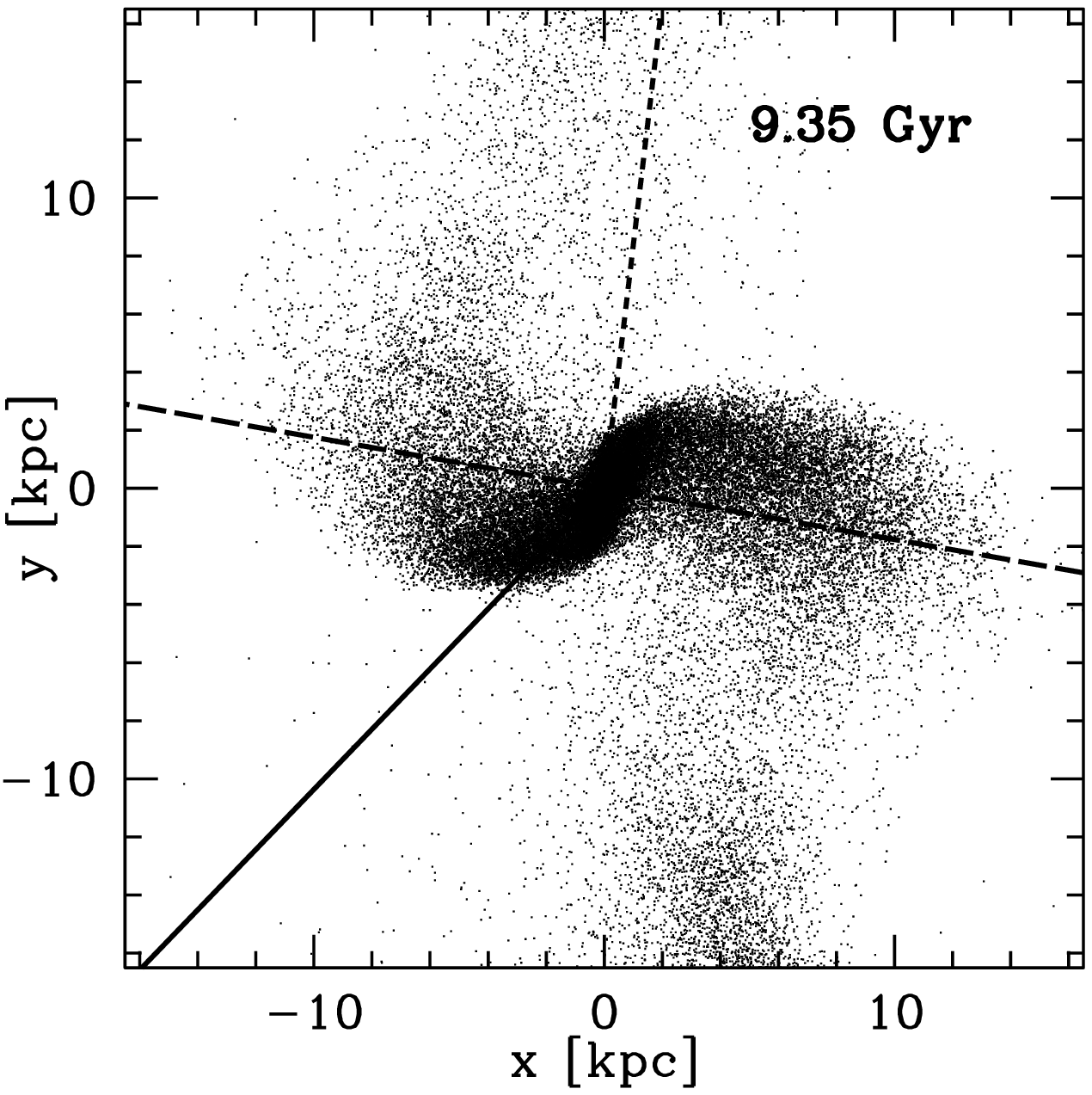}
    \epsfxsize=5.8cm
    \epsfbox[0 0 400 400]{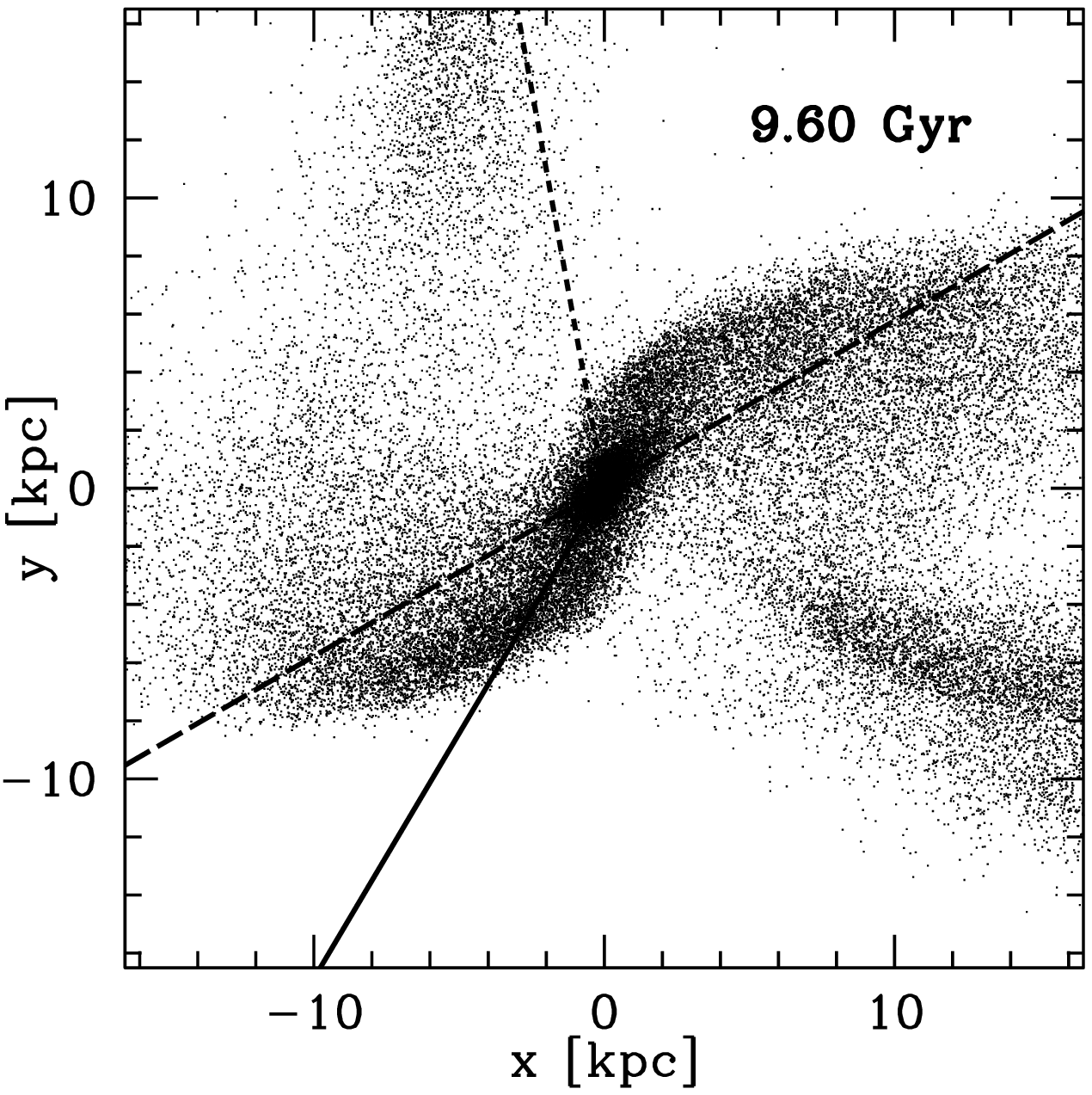}
\end{center}
\caption{Tidal tail flipping. The three panels present three snapshots from the simulation at
9.1, 9.35 and 9.6 Gyr from the start, corresponding to the positions on the orbit marked with open circles in
Fig.~\ref{orbit}. Dots show positions of 20 per cent of stellar particles
projected onto the orbital plane. The solid line indicates the direction to
the host galaxy, the short-dashed one is parallel to the velocity vector of the dwarf and the long-dashed one shows
the direction of the inner tidal tails. Old tidal tails oriented almost along the orbit decay and new ones pointing
almost radially towards the host galaxy are formed.}
\label{flipping}
\end{figure*}

Fig.~\ref{orbit} and the upper panel of Fig.~\ref{tails_properties} demonstrate
that while the transition from the radial to the
orbital (i.e. parallel to the orbit) orientation of the tidal tails happens rather smoothly
when the dwarf approaches pericentre, the transition from the orbital to the radial orientation
after a pericentre passage is rather abrupt. This is very well visible after the third, fourth and fifth
pericentre passage and happens roughly halfway between the pericentre and the apocentre
over a rather short timescale. Fig.~\ref{flipping} presents three
snapshots from the simulation showing this remarkable
transition starting at 9.1 Gyr after the beginning of the simulation, when the dwarf has almost reached its final
form. The dwarf has just passed the pericentre and its tidal tails are perpendicular to the radial direction
(towards the host galaxy). The tidal force acts perpendicular to the tails and dissolves them. This process is
accompanied by the stripping of a new material from the dwarf galaxy and formation of new, much denser and radially
aligned tidal tails. We propose to call this phenomenon the `tidal tail flipping'.

The orientation of the tails results from the interplay between the change
of orbital velocity and the direction of tidal force. For circular orbits the configuration does not change in time:
the velocity of the dwarf is always perpendicular to the direction to the MW and the tidal force acts always
perpendicular to the velocity. The orientation of the tails will therefore be the same in all parts of the orbit but it
will depend on the host potential and the mass of the dwarf. In the immediate vicinity of the dwarf the tails consist
of stars recently ejected from the dwarf so they point along the tidal force, that is along
the direction to the MW. As the ejected stars become unbound they will travel on their own orbits in the MW potential:
those that were ejected towards the MW will travel faster than the dwarf and form a leading tail, while those that were
ejected away from the MW will travel slower and form a trailing tail.

If the orbit is eccentric the situation becomes more complicated because the tidal force still acts towards the MW but
the velocity of the dwarf changes. At any moment the picture is similar to the case of circular orbit, but the
orientation of the tails will change in time, so it is better to speak about an inside and outside tail (with respect
to the dwarf's orbit). For example, when approaching the apocentre the unbound stars in the inside tail will reach
their apocentres sooner than the dwarf (they are on tighter orbits in the MW potential) and will slow down, while those
in the outside tail will not yet reach their apocentres and will move faster. The net effect we see is that near
apocentres the tails are oriented along the direction to the MW. On the other hand, when the dwarf is on the way to its
pericentre, the stars in the inside tail speed up with respect to the dwarf (those in the outside tail slow down) and
the tails become elongated along the orbit. After the pericentre the situation is not exactly symmetrical because of
the impulsive mass loss at that time. This new material is ejected mainly along the direction to the MW and the orbits
of these newly ejected stars behave as always when approaching the apocentre (the radial alignment of the tails is
preferred). The two factors combine so that the net effect is the formation of strong tails along the direction to the
MW that we have referred to as the `tidal tail flipping'.

\begin{figure}
\begin{center}
    \leavevmode
    \epsfxsize=7.5cm
    \epsfbox[0 0 370 370]{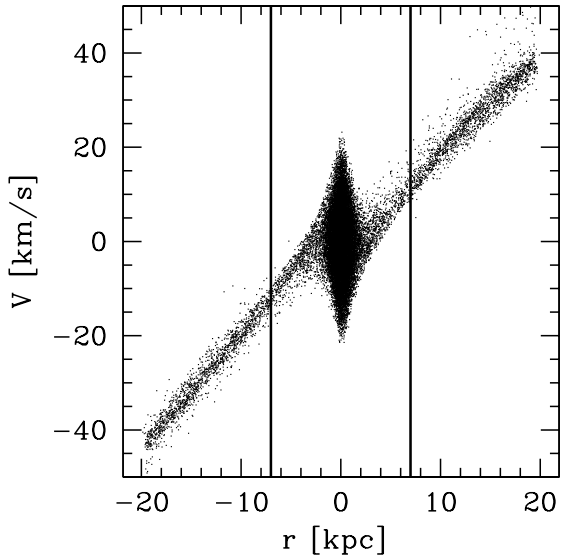}
\end{center}
\caption{Velocities of stellar particles of the simulated galaxy at $t=8.5$ Gyr
projected onto the axis parallel to the tidal tails
(and intersecting the centre of the dwarf galaxy). The coordinate
$r$ measures the distance from the centre of the dwarf projected onto this axis.
Vertical solid lines show the limits within which most stars are
bound. Only 15 per cent of the stars were plotted for
clarity.}
\label{veltail}
\end{figure}

\section{The kinematics of inner tidal tails}

The middle panel of Fig.~\ref{tails_properties} shows the relative
density of stars in the tidal tails at the distance of 10 kpc from the dwarf galaxy (dashed line)
compared to the density everywhere around the galaxy (measured in a ring placed in the orbital plane, solid line).
These measurements illustrate the mechanism for tail formation.
Most matter is stripped during the
pericentre passages but tidal tails enhance their density a little later along the orbit reaching the
maximum density typically about 0.5 Gyr after pericentre.
This happens because some time is required for stars to travel several kpc away
from the galaxy and form the tails. After reaching the maximum, the density drops exponentially
until the next pericentre passage, after which the previously formed tails are dissolved
and new tails are formed, as described in the previous section.

\begin{figure*}
\begin{center}
    \leavevmode
    \epsfxsize=5.8cm
    \epsfbox[0 0 400 400]{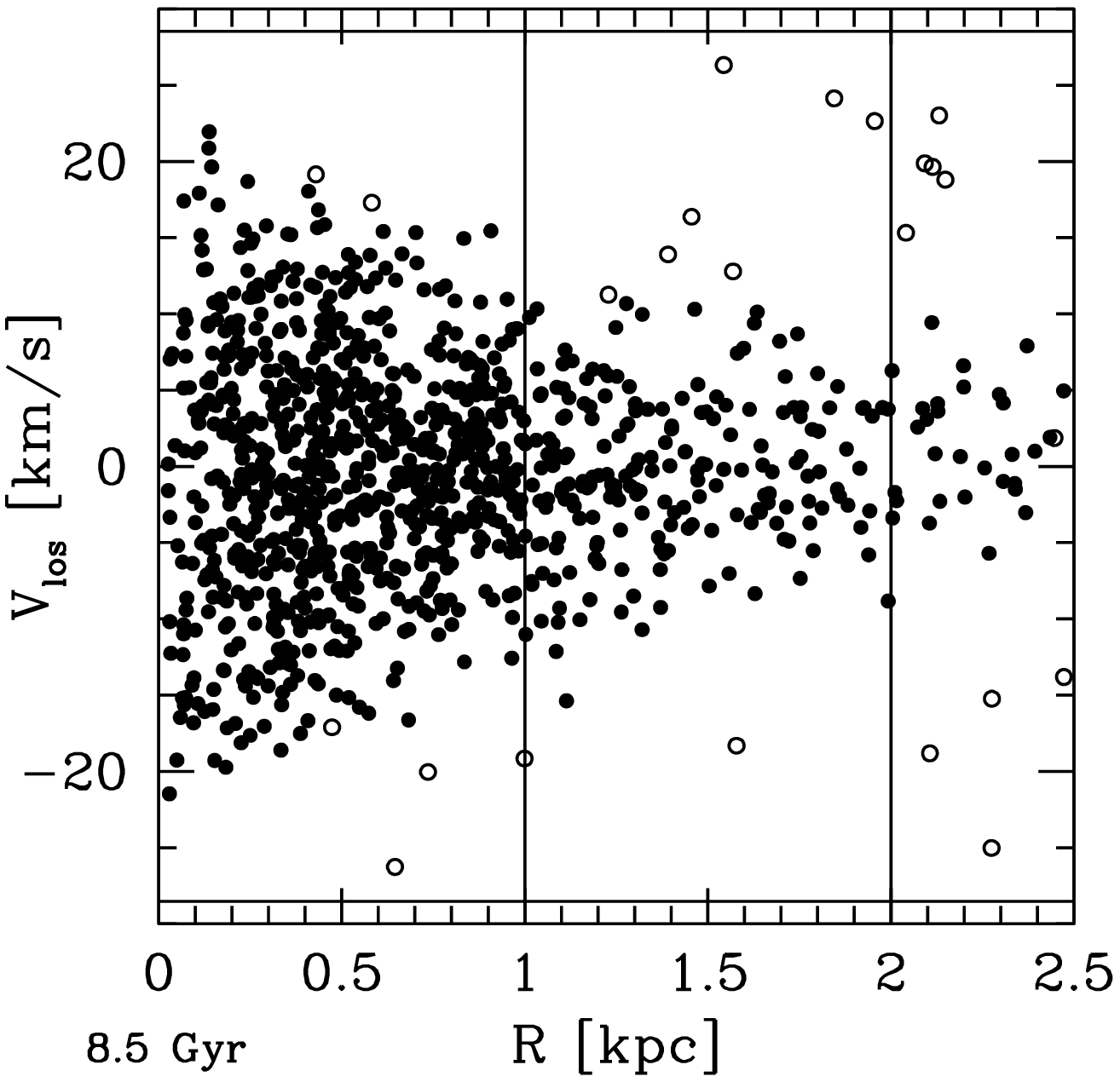}
    \epsfxsize=5.8cm
    \epsfbox[0 0 400 400]{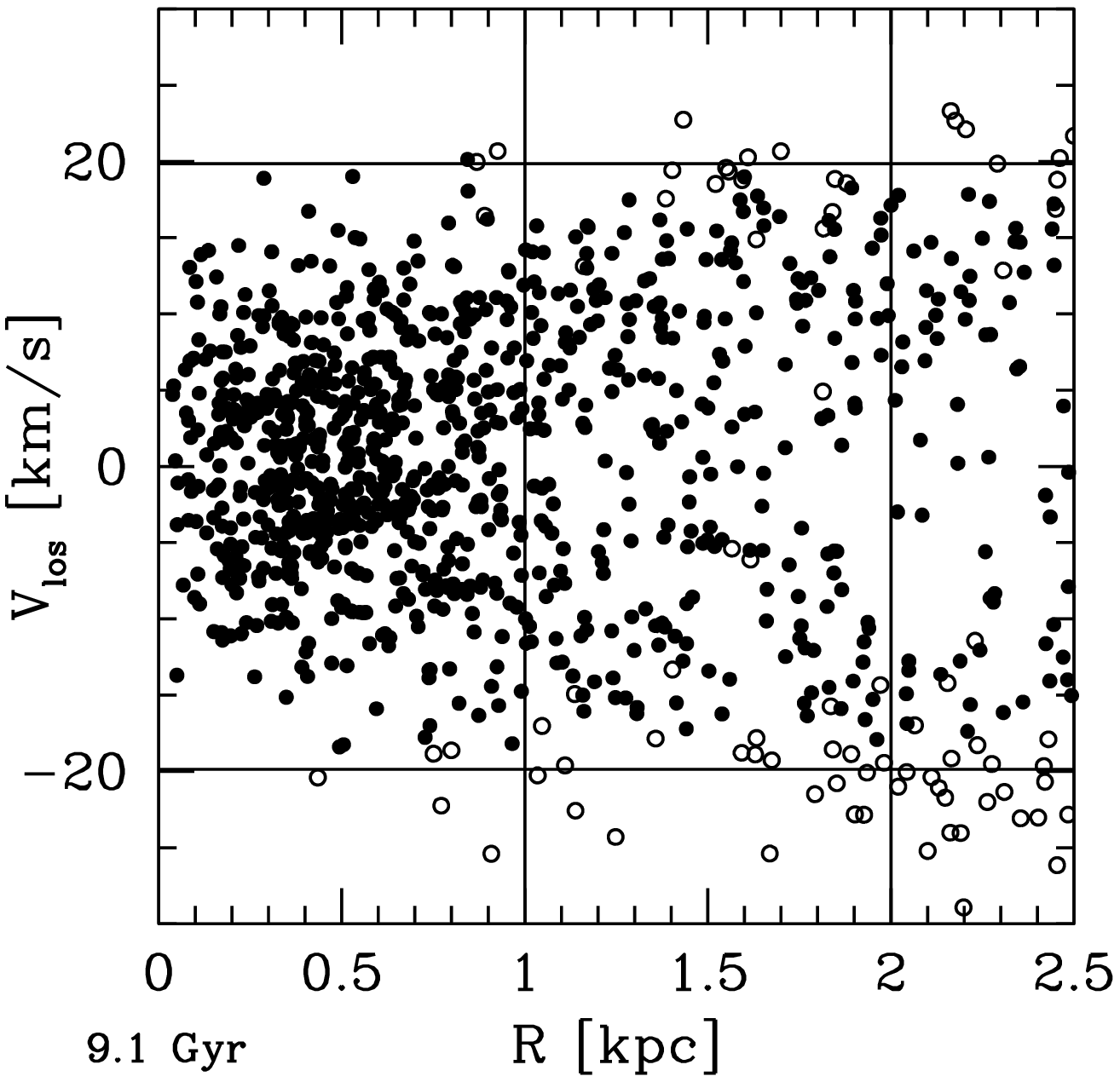}
    \epsfxsize=5.8cm
    \epsfbox[0 0 400 400]{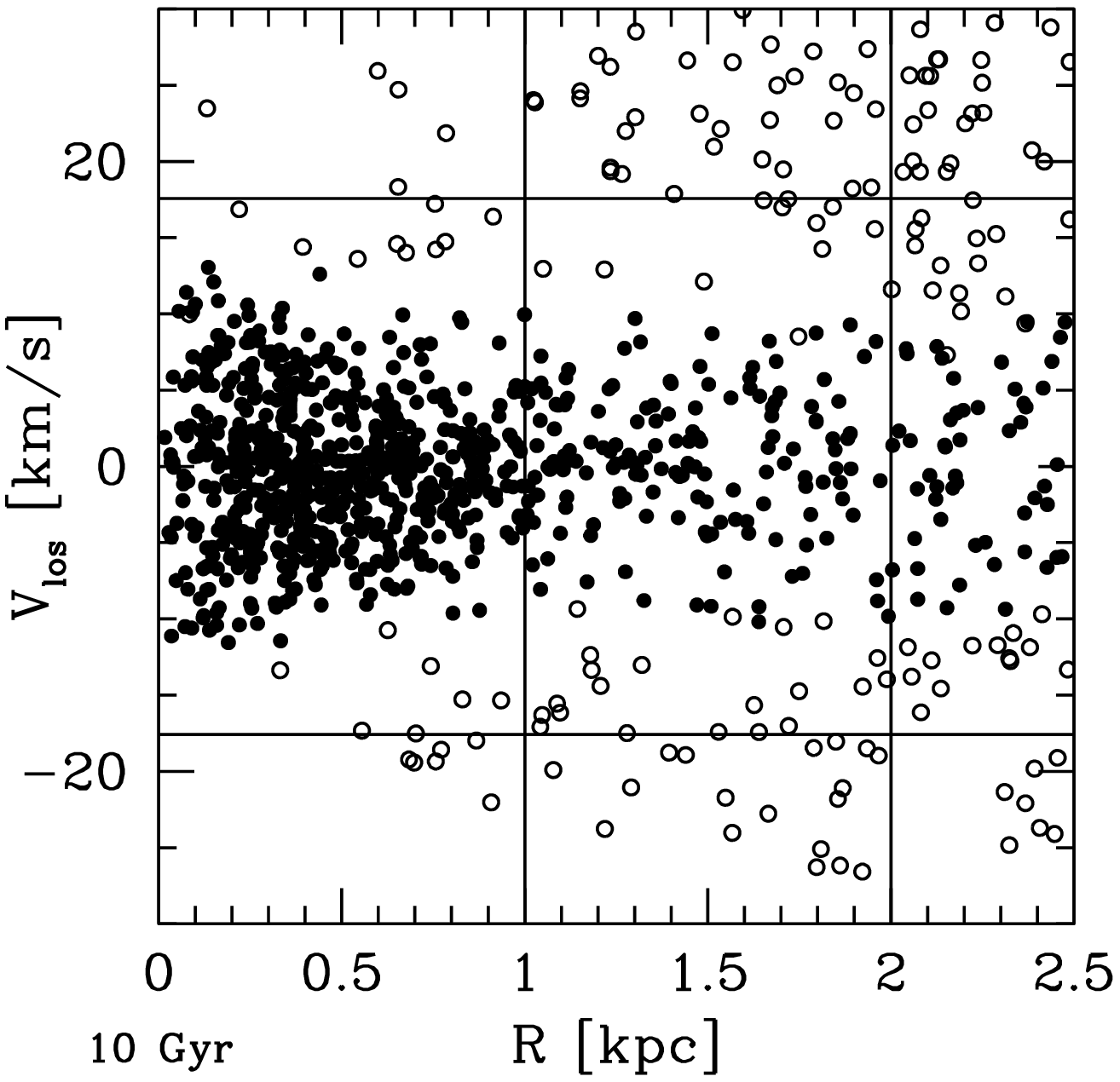}
\end{center}
\caption{Velocity diagrams of the stars at 8.5, 9.1 and 10 Gyr (from the left to the right panel). The plots show the
line-of-sight velocities of a thousand randomly selected stars with respect to the dwarf mean velocity as a function
of their projected distance from the centre of the dwarf. Filled circles indicate bound stars, open circles the unbound
ones. Horizontal lines show the velocity cut-off $|V_{\rm los} | < 3\sigma_0$ and the vertical lines the boundaries in
projected radius defining the kinematic subsamples with $R<1$ kpc and 1 kpc $< R <$ 2 kpc.}
\label{diagrams}
\end{figure*}

This picture suggests that it should be much easier to detect
tidal tails for dwarf galaxies shortly (few hundred Myr) after pericentre, i.e. when they have positive radial
velocities with respect to the host. The parts of the diagram shown in the middle
panel of Fig.~\ref{tails_properties} where the solid line is higher than the dashed indicate that the tidal tails are
then highly diffuse. There is always a
significant number of stars distributed uniformly around the galaxy but prominent tidal tails are not
always present as a dwarf satellite evolves in the primary's potential.

Fig.~\ref{veltail} plots velocities of the stellar particles inside the
dwarf galaxy and its tidal tails projected on an axis parallel to the tidal tails as a function
of the distance from the centre of the dwarf measured along the same axis. The dependence of velocity on distance is
a well known feature of tidal formations commonly referred to as the `velocity gradient' (e.g. Piatek \& Pryor
1995). Interestingly, we find that there is a tight linear relation between the radial velocity of a star and its
distance from the galaxy. It turns out that this relation is valid during most of the time except for the pericentre
passages. Only the slope of the $V(r)$ relation varies with time.

The stars enclosed between the vertical solid lines in Fig.~\ref{veltail} nearly all have
velocities lower than the escape velocity, thus they are still bound to the galaxy and
are expected to fall back, while stars outside will eventually feed the tidal streams.
The lower panel of Fig.~\ref{tails_properties} shows with dashed lines the values of
${\rm d}V/{\rm d} r$ (the slope of the relation) along with the distance of
the dwarf from the host galaxy (solid line) as a function of time. Clearly, the slope
of the relation is anticorrelated with the distance from the host galaxy.
After the pericentre passage, when the tails are created, the $V(r)$ relation
is steeper and it flattens with time. After the apocentre the relation becomes steeper
again. The numerical values plotted in the lower panel of Fig.~\ref{tails_properties} were
obtained by fitting a linear function to the velocity diagrams similar to the one
presented in Fig.~\ref{veltail} for every simulation output. The fitting was done
for stars within the distance range $5<|r|<20$ kpc from the centre of the dwarf.

It is important to emphasize that the presence of a star in the tidal tails
does not imply that it is unbound to the galaxy. Some of the
stars from the tidal tails are actually falling back onto the dwarf.
The bound region marked by the vertical solid lines in Fig.~\ref{veltail}
varies with time. When the dwarf travels from pericentre to apocentre unbound stars leave its vicinity quickly, while
the fastest bound stars reach their turnaround points shortly after the apocentre. This extended region, dominated by
bound stars, can reach up to 10 kpc from the galaxy.

\begin{figure}
\begin{center}
    \leavevmode
    \epsfxsize=8.5cm
    \epsfbox[20 0 530 185]{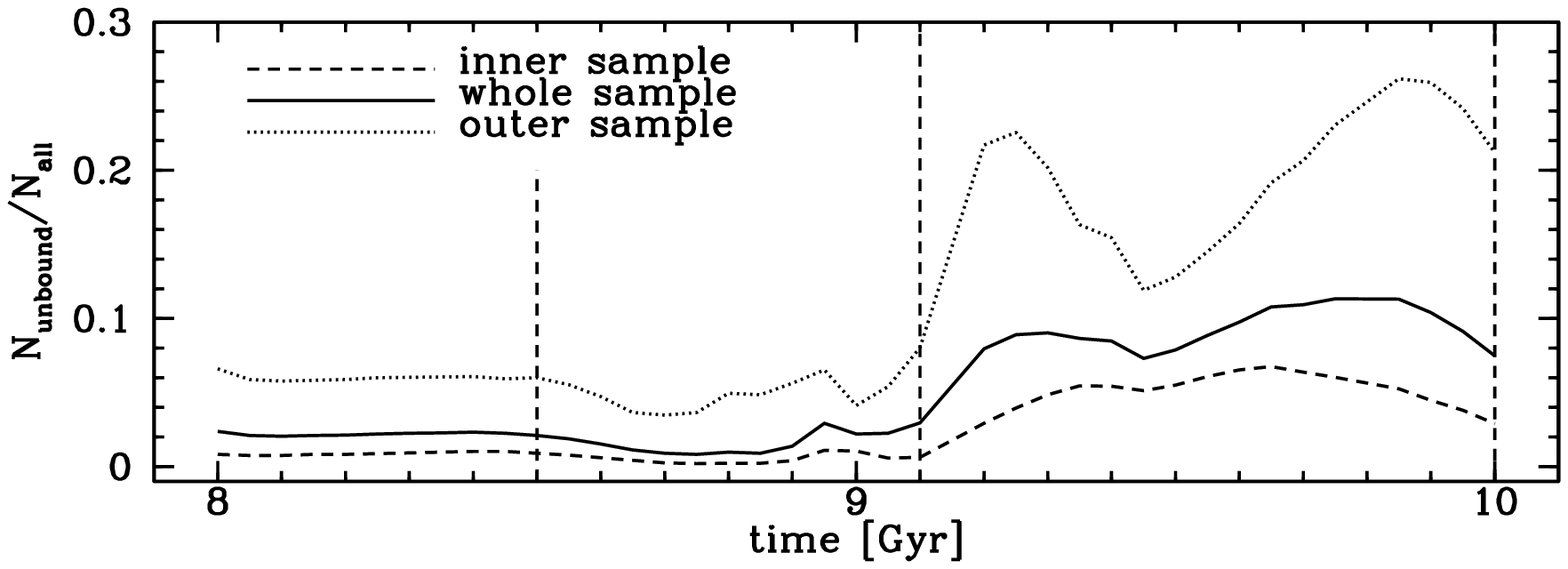}
    \leavevmode
    \epsfxsize=8.5cm
    \epsfbox[20 0 530 185]{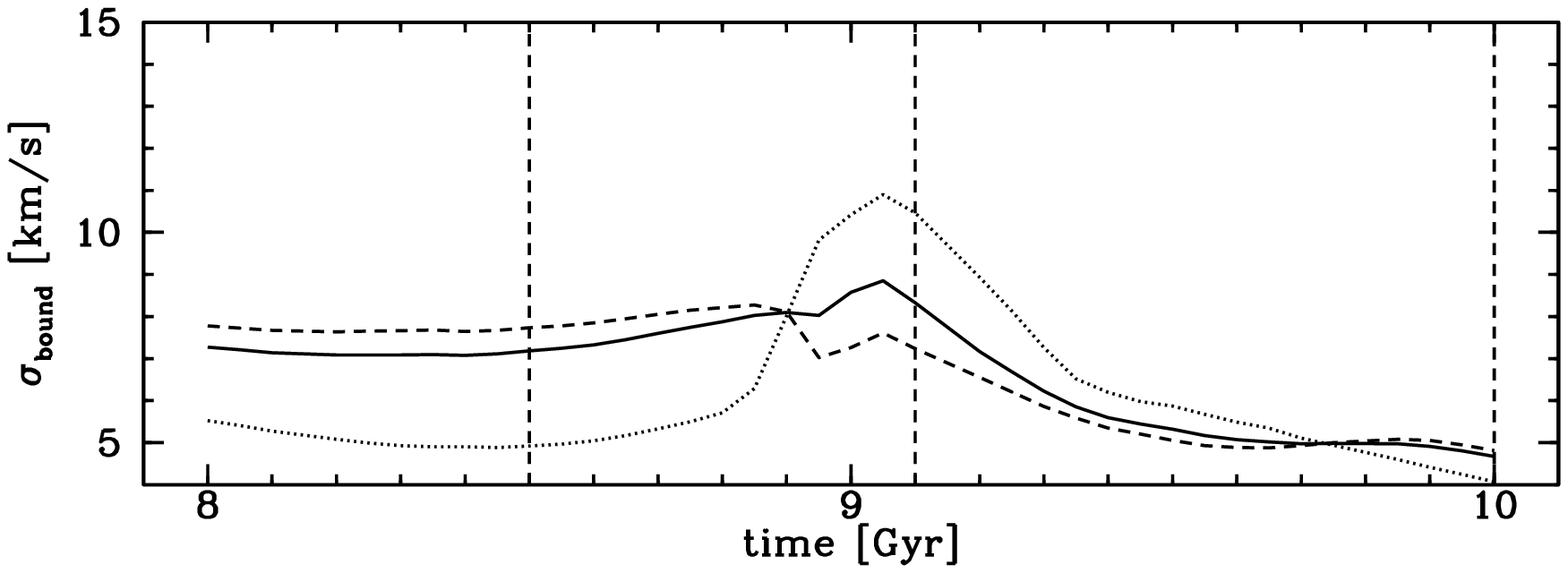}
    \leavevmode
    \epsfxsize=8.5cm
    \epsfbox[20 0 530 185]{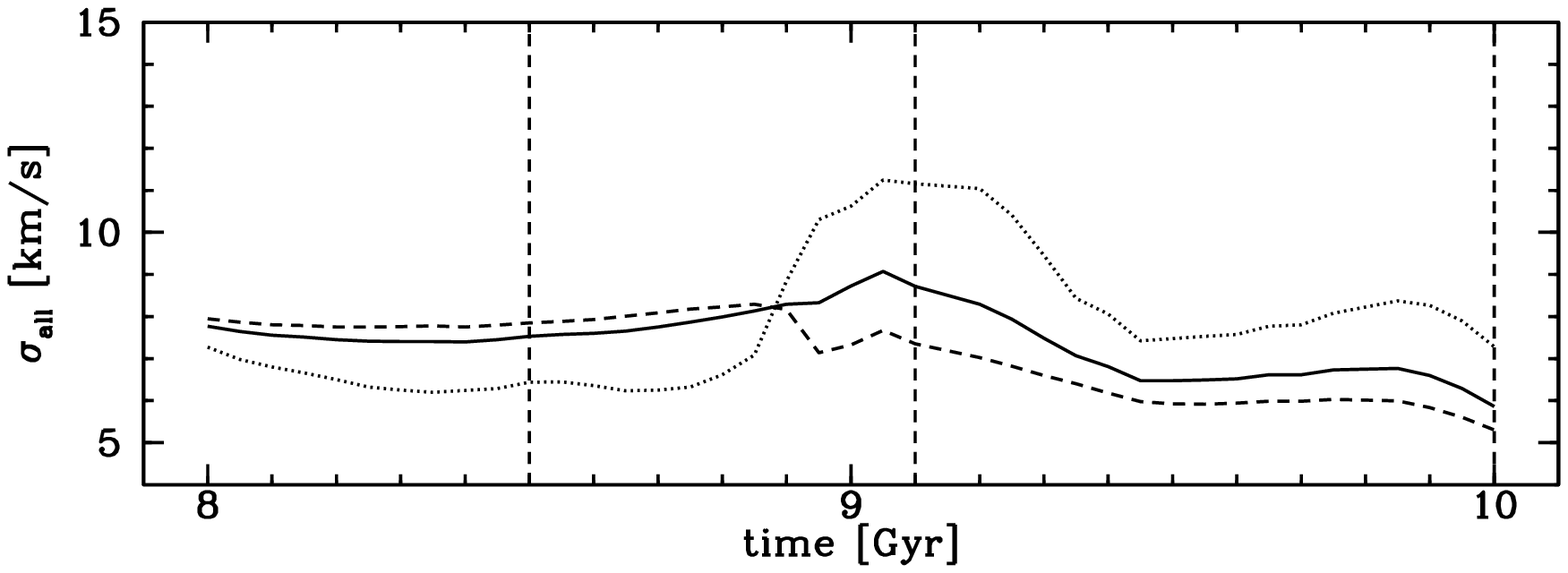}
    \leavevmode
    \epsfxsize=8.5cm
    \epsfbox[20 0 530 185]{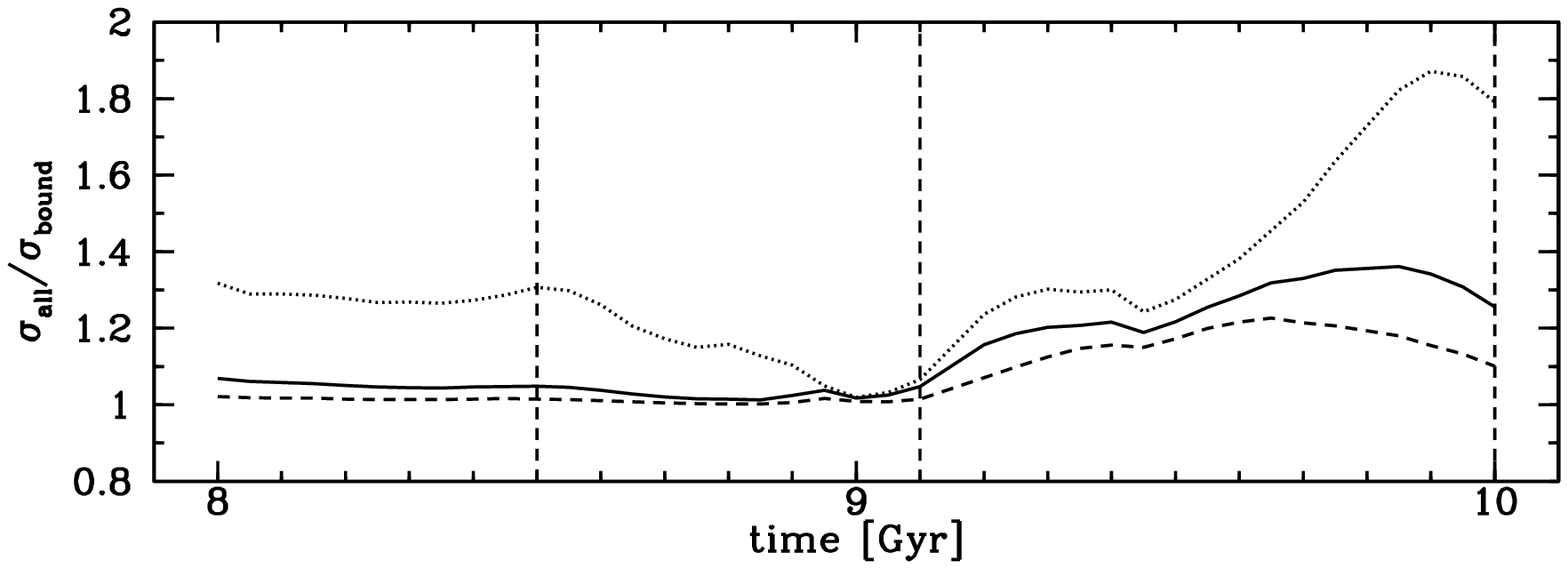}
\end{center}
\caption{The effect of tidal tails on the measured velocity dispersion of the dwarf as a function of time over the last
orbit. All panels refer to the kinematic sample obtained by selecting stars within a given range of projected radii $R$
and introducing the velocity cut-off $|V_{\rm los} | < 3 \sigma_0$. In each panel the solid line corresponds to all
stars within $R<2$ kpc, the dashed line to the inner sample with $R<1$ kpc and the dotted line to the outer sample with
1 kpc $< R <$ 2 kpc. The upper panel shows the fraction of unbound stars in each sample. The second and third panels
plot the velocity dispersion of the bound and all stars respectively and the fourth panel shows the ratio of the two.
Vertical dashed lines indicate snapshots for which the velocity diagrams in Fig.~\ref{diagrams} are shown.}
\label{contamination}
\end{figure}

\section{The contamination of kinematic samples}

In this section we address the question of how the kinematic samples and measured velocity dispersions are affected by
the presence and properties of tidal tails in the immediate vicinity of a dwarf galaxy. The problem has been already
discussed in detail by Klimentowski et al. (2007) using a single, final output of a similar simulation. Here we
consider a similar problem but taking into account the actual variation of the orientation and density of the tails as
the dwarf travels on its orbit around the host galaxy. We restrict the analysis to the last full orbit, i.e. the time
between the two last apocentres, between 8 and 10 Gyr from the start of the simulation. This time is
representative so that all configurations of the tails are covered and on the other hand the dwarf is already evolved
enough so that it can actually be described as a dSph galaxy (see Klimentowski et al. 2009 and {\L}okas et al. 2009
for details).

The kinematic samples are constructed in a similar way as in real observations. We place an observer in the centre of
the MW and project the velocities of the stars in the dwarf along the line of sight and their positions on the surface
of the sky. Three examples of velocity diagrams constructed in this way are shown in Fig.~\ref{diagrams} for the
outputs at 8.5, 9.1 and 10 Gyr from the start of the simulation. The diagrams show the line-of-sight velocities of the
stars versus their projected radii. To select the stars we have applied a rather conservative cut-off in velocity
$|V_{\rm los} | < 30$ km s$^{-1}$ with respect to the dwarf's mean and the projected radius $R<2.5$ kpc. The stars
bound to the dwarf are plotted with filled circles, those unbound with open circles.

At 8.5 and 10 Gyr (the left and
right panel of Fig.~\ref{diagrams}), where the dwarf is close to apocentres, the velocity diagrams are regular, with a
characteristic triangular shape typical of relaxed objects. At 9.1 Gyr (middle panel) the dwarf has just passed its
pericentre and is strongly perturbed. The forming tidal tails are visible at larger $R$ as two separate branches. Note
that although the stars clearly belong to the tails, very few of them are actually already unbound, but they are in the
process of being stripped and cannot be considered as reliable tracers of the dwarf potential. The unusual shape of the
velocity diagram in this case could be used as a warning against using such samples for dynamical modelling.

For the further analysis we restricted the sample to the projected radii $R<2$ kpc and introduced a cut-off in the
line-of-sight velocity $|V_{\rm los} | < 3\sigma_0$ with respect to the dwarf's mean, where $\sigma_0$ is the central
velocity dispersion measured in each simulation output for stars within $R<0.2$ kpc where the contamination is very low.
This is the standard procedure used in selecting kinematic samples for dynamical analysis, based on the assumption that
the velocity distribution is approximately Gaussian. We adopted a variable velocity cut-off because the bound mass of
the dwarf changes between 8 and 10 Gyr by more than a factor of two, from $8 \times 10^7$ M$_{\sun}$ to $3 \times
10^7$ M$_{\sun}$ due to tidal stripping (see Klimentowski et al. 2009). The radius $R=2$ kpc corresponds to
the size of the dwarf at the final stage, i.e. at larger radii the tidal tails start to dominate.

For kinematic samples constructed in this way we measured the fraction of unbound stars in the total sample, the
line-of-sight velocity dispersion of bound and all stars and the ratio of the two. Velocity dispersions were calculated
using the standard unbiased estimator (see e.g. {\L}okas, Mamon \& Prada 2005). We made separate measurements for the
whole sample with $R<2$ kpc, the stars with radii $R<1$ kpc (inner sample) and 1 kpc $< R <$ 2 kpc (outer sample). The
results are illustrated in the four panels of Fig.~\ref{contamination} with the solid line corresponding to the whole
sample, the dashed one to the inner sample and the dotted one to the outer sample.

We immediately see that the contamination is
rather different before and after the pericentre passage. After the pericentre the fraction of unbound stars is much
higher (it reaches 26 per cent for the outer sample) and has two maxima, the first one due to the formation of new
strong tidal tails and the second due to the tails becoming more radially oriented. The velocity dispersions are
obviously most affected at the pericentre when the dwarf is most strongly perturbed, even if its tidal tails are not
oriented along the line of sight. As already mentioned, the stars which are being ejected at this moment, although
formally still bound to the dwarf cannot be reliably used for dynamical modelling. After the pericentre, the presence
of the tails can increase the central velocity dispersion by up to 20 per cent, the total velocity dispersion by up to
40 per cent, but the dispersion of the outer sample can be inflated by almost a factor of two.

An obvious remedy to this effect would be to use for dynamical modelling only the inner sample where the
contamination is relatively low. Note however, that with growing kinematic samples presently available for dSph
galaxies restricting the analysis to the innermost part of the dwarf would mean to lose a lot of information. It is now
common in such analyzes to use velocity dispersion profiles rather than single dispersion values since only the profiles
allow us to constrain the actual distribution of mass in the dwarfs rather than just the total mass. Without this,
addressing such issues as the shape of dark matter profile or the cusp/core problem would be impossible. In addition,
as demonstrated by Klimentowski et al. (2007) such contamination can be effectively removed by common procedures
for interloper rejection.

\section{Discussion}

We studied the properties of tidal tails in the immediate vicinity of dwarf galaxies orbiting in a MW potential.
We find that for most of the time tidal tails point almost radially towards the host galaxy. This behaviour
is in agreement with Montuori et al. (2007) who found the same effect in their
$N$-body simulations of globular clusters. Therefore the orientation of tidal tails in the immediate vicinity of the
galaxy is not directly related to the direction of the orbital motion. Only unbound stars lost much earlier and forming
large-scale extensions further out from the bound component of the dwarf tend to follow the orbit.

In contrast, Combes, Leon \& Meylan (1999) and Dehnen et al. (2004) find that the tidal
tails in their globular cluster simulations follow the orbit. The difference is mainly due to the fact that they study
tails out to much larger distances than we did here. Some differences are also expected because of the different orbits
which they assumed to be less eccentric than in our case. Johnston et al. (2002) have shown that indeed for almost
circular orbits tidal tails are pointed radially towards the host galaxy only very close to the cluster. Other
differences come from the fact that globular cluster simulations do not involve dark matter and all simulations
performed to date started with a spherical object while our dwarf initially possessed a stellar disk.

The radial alignment of tidal tails may at least partially explain the difficulty
of detecting them by photometric observations. Since we are looking at the dwarfs from
the inside of the MW their tidal tails are most probably oriented along the line of
sight and they are likely to be confused with a much brighter dwarf galaxy itself.
In the rare cases when the line of sight is perpendicular to the tidal tails the tails will
generate an excess of star
counts at a few core radii from the centre, which will show up as a flattening of the
outer part of the star counts profile (Klimentowski et al. 2007; Pe\~narrubia et al. 2009). Such a flattening might
have been observed in some dSphs but not in all of them (Mu\~noz et al. 2006; Mu\~noz, Majewski \& Johnston 2008).

The most important consequence of the radial alignment of tails is the fact that the tails are aligned with the
line of sight of the observer who is placed inside the host galaxy. This means that even far away from pericentre, when
the dwarf appears unperturbed and relaxed, the kinematic samples of stars will be
contaminated by unbound stars from the tails. This contamination should be highest for dwarfs receding from the MW
right after the tidal tail flipping. Given the tight and steep relation between the distance of stars in the tails and
their velocity this means that even the relatively near tidally stripped stars must artificially inflate the measured
velocity dispersion, especially in the outer regions of the dwarfs. It is therefore essential to remove these unbound
stars in order to obtain reliable mass estimates for dwarf galaxies.

Klimentowski et al. (2007) showed that this can be very
effectively done using a procedure proposed by den Hartog \& Katgert (1996). This procedure has been recently applied
by {\L}okas (2009) to new, large kinematic samples for the Fornax, Carina, Sculptor and Sextans dwarfs published by
Walker, Mateo \& Olszewski (2009). It has been shown there that the contamination is indeed significant and once it is
removed the velocity moments are well reproduced by solutions of the Jeans equations for models where mass follows
light, as predicted by the tidal stirring scenario (Klimentowski et al. 2007, 2009; Mayer et al. 2007; {\L}okas et al.
2008, 2009).

\section*{Acknowledgments}

SK is funded by the Center for
Cosmology and Astro-Particle Physics (CCAPP) at The Ohio State University.
JK is grateful for the hospitality of CCAPP during his visit.
The numerical simulations were performed on the
zBox2 supercomputer at the University of Z\"urich.
This research was partially supported by the
Polish Ministry of Science and Higher Education
under grant NN203025333 and the LEA
Astronomie France Pologne program of CNRS/PAN.
We also acknowledge the LEA program and the ASTROSIM network of the European Science Foundation (Science
Meeting 2387) for the financial support of the workshop `The local universe: from dwarf galaxies to galaxy clusters'
held in Jablonna near Warsaw in June/July 2009, where part of this work was done.

\end{document}